\documentclass[journal]{IEEEtran}
\usepackage{blindtext}
\usepackage{graphicx}
\usepackage{lettrine}
\usepackage{amsmath}
\usepackage{textcomp}
\newcommand\addtag{\refstepcounter{equation}\tag{\theequation}}
\usepackage{caption}
\usepackage{hyperref}
\usepackage[all]{hypcap} 

\ifCLASSINFOpdf

\else

\fi
\begin{document}

\title{Ultra Low-Power System for Remote ECG Monitoring}

\author{
        Ehsan~Hadizadeh,
        Rozhan~Rabbani,
        Zohreh~Azizi,
        Matin~Barekatain,
        Kourosh~Hakhamaneshi,
        Erfan~Khoram,
         and Ali~Fotowat-Ahmady
\thanks{E. Hadizadeh, R. Rabbani, and Z. Azizi are with the Department of Electrical Engineering, Sharif University of Technology, Tehran, Iran (e-mail: \href{mailto:Hadizadeh_e@ee.sharif.edu}{hadizadeh\underline{\hspace{.05in}}e@ee.sharif.edu}; \href{mailto:rabbani_rozhan@ee.sharif.edu}{rabbani\underline{\hspace{.05in}}rozhan@ee.sharif.edu}; \href{mailto:azizi_zohreh@ee.sharif.edu}{azizi\underline{\hspace{.05in}}zohreh@ee.sharif.edu}).}}

\maketitle

\begin{abstract}
A complete system solution extracting signals from the patient chest with three leads including motion artifact removal in both analog and digital implementations are described. The resulting ECG signal is transferred via Bluetooth low energy to a mobile phone. Using deep sleep modes, the overall power consumption is less than 300$\mu$A and the device can operate for more than 20 days using a 150mAh battery. The screening software looks for suspicious traces such as those with missing pulses, tachycardia, bradycardia, etc. The mobile phone software also eliminates any remaining motion artifact. The traces are subsequently processed in detail in a cloud server and to a physician’s dashboard for long-term monitoring.
\end{abstract}


\IEEEpeerreviewmaketitle

\section{Introduction}

\lettrine[findent=1pt]{\textbf{A}}ccording to the latest releases of World Health Organization (WHO) the foremost reasons of death all over the world are cardiovascular diseases (CVDs). In 2015, approximately  17.7 million people died from CVDs accounting for 31\% of all deaths around the world. 7.4 million of these deaths were caused by coronary heart disease and 6.7 million were due to stroke. 

Life-threatening behaviors such as consumption of tobacco products, unhealthy diet, sedentary lifestyle and obesity, and deleterious use of alcohol should be regarded to prevent most cardiovascular diseases. Early detection and treatment are necessary for people suffering from cardiovascular diseases or who are at high cardiovascular risk. \cite{WHO}
  
Today, wearable sensors play a prominent role in health-care products. The most significant reason is that they facilitate continuous real-time monitoring of a lot of people concurrently. Added, they pave the way for self-monitoring of vital signals such as heart signal. The continuous observation of heart signal leads to an early detection of fatal risks. 

Demographic trends exhibit that by 2050, the world population with more than 65 years will be more than the world population of age 15 and less \cite{Jsenone}. This elderly population are greatly susceptible to the risk of chronic diseases such as CVDs. It is estimated that broad adoption of remote monitoring technologies for chronic diseases saves a net annual of 12 billion dollars in us health care expenditure. So added to the advantage of early detection of lethal risks, application of wearable biomedical devices causes remarkable economic advantage by means of reduction in physician and emergency room visits, hospitalization and nursing home care. \cite{Jsenfour} 

Also a giant market sector for wearable biomedical devices are associated with a remarkable portion of physically active and health-conscious population who are concerned about tracking their well-being. And between the two extremes, there lies a huge population of diabetic, obese, elderly people taking the most advantage of services and devices providing them with a better health self-management. \cite{Jsenfive}

Due to the paramount importance of wearable biomedical devices, researchers have tried to elevate the effectiveness of wearable systems. Improving the accuracy and resolving the issues related to such systems should be regarded as one of scientists top priorities as the demand for wearable sensors increases day by day. 

So In this manuscript, we present a wearable system for multi-person remote long-term ECG monitoring in order to acquire continuous real-time data about users' health status in residential settings without any need for others' help. The emphasis of our work is on lowest power consumption possible. Motion artifact removal (MAR) is highly considered in our work due to a couple of reasons. First, motion artifact can lead to a low quality signal which causes deficient clinical diagnosis.\cite{Tbcas} Second, motion artifact can be misinterpreted with event signals that the device is deliberately sensitive to. This may result in false event detections and alarms \cite{JsenMAR}. Third, motion artifact can saturate the analog circuit which leads to wrong results . The high motion artifact removal of this work extends its applicability to everyday-life situations \cite{JsenMAR}. Also, it has a robust operation in everyday activities and is easy to use. This work is designed for a variety of applications, ranging from computing heart rate to sophisticated high-accuracy biomedical diagnostics under arbitrary conditions.

\begin{figure}[h]
    \centering  
    \includegraphics[width=0.45\textwidth]{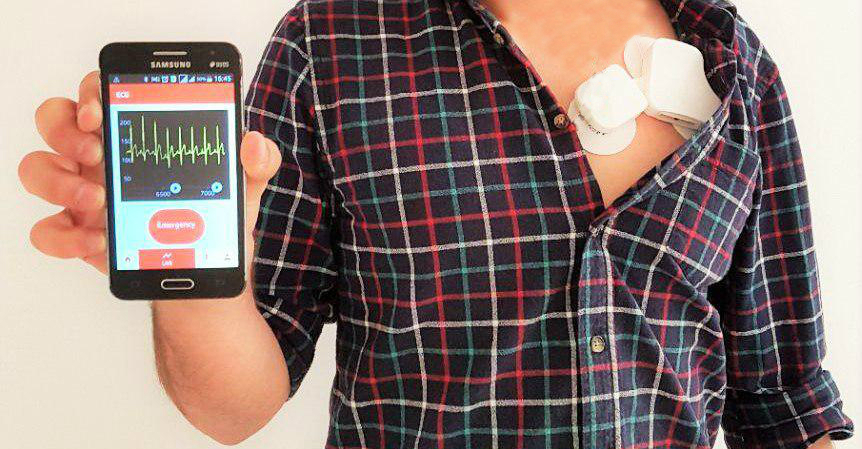}
    \captionsetup{justification=centering}
    \centering {\caption{\label{fig:1} The proposed ultra-low power system of ECG remote monitoring}}

\end{figure}

The system shown in Fig. \ref{fig:1} includes a wearable Ultra-low power hardware which is battery-based. The hardware\textquotesingle s main function is data acquisition from the ECG sensors and a simultaneous data transfer. Obviously, it is not possible to save all data on the memory of the mentioned hardware because of its small size and its low power usage. So the data transmission to a smartphone is required. The software part of the system involves three user-friendly applications. First on smartphone, second on patient\textquotesingle s personal computer and third a dashboard for the user\textquotesingle s physician. As these three applications are connected synchronously and by cloud, they provide a true “real-time” health monitoring system both for the patient and the corresponding physician.

There are some methods by which data can be sent from the hardware to a smartphone such as Wi-Fi, ZigBee, Bluetooth and BLE. While there is a trade-off between the size of the battery and its active duration in the circuit, it is of great significance to reduce the average power consumption of the whole hardware. There is a strong conflict between reducing the average power consumption and using Wi-Fi for transmission. ZigBee is not supported in smartphones which play a key role in our system. Moreover, local data transmission in our work eliminates the need to use a communication protocol like ZigBee which provides a wide range of data transmission with the cost of reducing data rate and limiting access to a variety of devices. Thus, a comparison between BLE and Bluetooth is a key factor in helping us to select the final method for data transmission. 

Normal Bluetooth is a way of data transmission by which it is possible to transfer data with the rate of 2Mbps. One of the main drawbacks of this approach is its energy usage diagram which is in the active region all the time. As a result, controlling the power consumption is not a facile process. That is why using normal Bluetooth method for data transfer is not desirable to be applied in this project. Pursuing all of the available methods, we infer that applying BLE method is the most suitable way for this ultra-low power system.

\section{System Overview}

\begin{figure}[h]
    \centering   
    \includegraphics[width=0.45\textwidth]{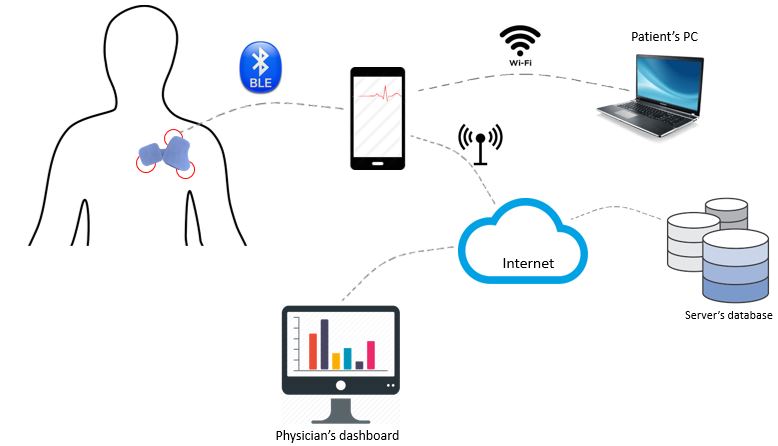}
    \captionsetup{justification=centering}
    \centering {\caption{\label{fig:2}Block diagram of the system}}

\end{figure}

The block diagram of the system is shown in Fig. \ref{fig:2}. The system has three main parts. Here each part\textquotesingle s main aspects are explained briefly.

\subsection{Sensor, Receiver \& Transmitter}

A light wearable hardware collects real-time data and sends it via BLE to the smartphone. In this stage, data packets, arranged by the co-processor, are sent using the main processor during active intervals. The data is received on the smartphone and is stored on its memory.

\subsection{High Performance Server}

By connecting the smartphone to the high performance server, data is shared and the physician can also have access to this data. DSP on ECG signal will be available on this server. Many functions such as determining fault in connection of the sensors to the body, detecting alarms related to heart disease, or identifying date of measurement can be processed using this server. 
The user\textquotesingle s personal computer can also connect to this server and transfer processed data whenever they are required. In this stage, user interface is also designed.

\subsection{User Interface}

Designing a user-friendly interface can boost the level of satisfaction of different users such as physicians, patients, athletes, etc. The interface contains three parts; smartphone, personal computer and physician\textquotesingle s dashboard. They provide the patients and the physicians with the advantage of storing, visualizing and sharing data in an efficient way. 

\section{System Design Considerations}

The hardware contains two main circuits as the analog and digital parts, which together with three electrodes extract ECG signal and pre-process it to get ready to be transmitted by Bluetooth Low Energy protocol.
\newline The user-friendly structure of the device with its rechargeable battery has made self-monitoring of ECG possible during daily activities. Pre-process of ECG signals is carried out by various parts, some of which are the Analog Front-End, the ADC and the Digital Processor. 

\subsection{Signal Characteristics \& Considerations}

The basics of ECG measurement are the same for all ECG applications ranging from monitoring heart rate to the diagnosis of specific heart conditions, but the requirements for electrical components vary greatly based on each application \cite{JsenTwentyEight}. The analog front end (AFE), as the part which extracts the ECG signal and prepares it for the digital process, should be designed on the basis of ECG signal characteristics and application requirements. The electrocardiogram (ECG) is the graphical recording of the electrical activity of heart \cite{ProperSkin}. Heart signals are picked up through electrodes connected externally to specific locations on body. The differential voltages between two of the electrodes are commonly referred to as “Leads” which defer in number from one to a maximum of twelve based on the application \cite{JsenTwentyEight}. In this heart sensor a single lead AFE is adequate for self-monitoring of ECG. The ECG signal has an actual bandwidth from 0.5Hz to about 150Hz. It has a peak-to-peak amplitude of approximately 1 mV but can reach 3 mV \cite{Jsen}.

The main components of the ECG signal are: The differential ECG signal, the common-mode signal, and the differential ECG offset \cite{ECGcom}.

The quality of the ECG signal is often corrupted due to several sources of noise. Respiration activities and  motions of the patient\textquotesingle s muscles which are called “Motion Artifacts” together with the motions of the instruments, such as loose skin-electrode contacts can cause baseline wander adding a DC offset of approximately 300 mV to the main ECG signal which makes the signal ramble \cite{ECGcom}. The baseline wander can affect quality of the signal and data analysis and may cause saturation in AFE blocks when it exceeds the circuit\textquotesingle s dynamic range. 
In addition to the low frequency baseline wander, high frequency noise caused by power line interferences, other electro-magnetic (EM) sources, and EM fields such as RF signals can cause corruption right in the ECG signal\textquotesingle s bandwidth which can saturate the analog circuit \cite{ReviewJsen}. The main advantage of using the third electrode in this design is that it couples the circuit ground to the ground of the body in order to reduce common-mode noises before they can saturate the amplifiers. Coupling the third electrode to the patient\textquotesingle s body causes the supply lines of the device to have the same common-mode signal as the signal paths so no common-mode noise will be amplified through the AFE circuitry.  In order to further suppress common-mode signals, instrumentation amplifier circuits with high common-mode rejection ratio (CMRR) are used in the analog front end. Another practical method to reduce common-mode signals, called DRL (Driven Right Leg) is to feed the body with the common-mode signal extracted from the initial ECG signal considering the appropriate sign and magnitude. Applying DRL necessitates wiring to the leg of the patient which limits wearability and convenience of the device. In our case, the desired CMRR is met by the previous methods and since wearability is an important concern in the whole design; therefore, in order not to consume extra power, no DRL circuit is implemented in the AFE. 

In what follows, some design considerations regarding BLE application and the ECG data acquisition are pointed out. 

\subsection{The BLE Application Considerations}

The average power of the transmitter as a function of the packet interval is \cite{Chandra}:

\[P_{avg} = \frac{P_{active} . T_{packet} + P_{leakage} . T_{interval}}{T_{packet} + T_{interval}} \addtag\]

In \cite{Chandra} Chandrakasan mentioned that in the limit of ultra-low duty cycle operations, with $T_{interval}\to\infty$ \[\lim_{T_{interval}\to\infty} P_{avg}=P_{leakage}\addtag\]

Although it seems to be a correct expression at first glance, the supervision timeout interval of the mentioned BLE connection should be considered. A fundamental characteristic of the BLE application is its time setting. There are various intervals and time settings with different concepts listed below:

\subsubsection{\textbf{Minimum \& Maximum Connection Interval}} These are 2-byte parameters that denote the minimum and maximum permissible connection time.\cite{Cypress}

\subsubsection{\textbf{Slave Latency}} This is a 2-byte value and defines the latency between consecutive connection events.\cite{Cypress}

\subsubsection{\textbf{Connection Supervision Timeout Multiplier}} This is a 2-byte value that denotes the low energy link supervision timeout interval. It defines the timeout duration for which a low energy link needs to be sustained in case of no response from the peer device over the low energy link.\cite{Cypress}

According to these definitions, we are not allowed to assume $\lim_{T_{interval}\to\infty} P_{avg}=P_{leakage}$ even in ultra-low duty cycle operations due to the fact that the connection supervision timeout stops the intervals at most after 9 connection intervals when the connection intervals approach their maximum value.

\subsection{High Resolution ECG}

For rebuilding a high resolution ECG signal, data should be sampled at no less than 1 Ksps \cite{highres}. In this project, each data packet contains of 20 bytes of ECG data. Because of the limitations of the BLE IC and the receiver of the smartphone, it is not feasible to send more than 5 packets during an active period of BLE. On the other hand, our goal is to receive at least 1000 packets per second. Hence, the operation\textquotesingle s duty cycle should be about 100ms.

In spite of the fact that the connection supervision timeout stops the inactivity of the BLE transmitter after about 32 seconds in the worst case, it is still necessary to optimize the power consumption of the BLE transmitter and other blocks in the sleep intervals ($T_{leakage}$) \cite{DatasheetCC}. The first idea crossing the mind at first glance is to optimize the power consuming blocks during sleep intervals. In what follows, we will explicate the effect of each block and show how to optimize the power  consumption of the whole system.

\subsection{Power Consuming Blocks}

To design an ultra-low power circuit, it is necessary to analyze which blocks cause energy consumption and control them in a way that they use as little energy as possible. By designing such a system, we prevent the domination of the power consumption of other blocks during the sleep and the active intervals over that of the BLE transmitter. To shed light on this issue, some of the main energy consuming blocks in this system are highlighted below:

\subsubsection{\textbf{The Analog Front-End}} In many applications, this block, which is always active, consumes a huge amount of energy especially when the power consumption is not a key factor for the design. In the ultra-low power system of ECG remote monitoring, one of the main specification of the design is to minimize the power consumption of the analog front-end block. Since the components of this block are off-the-shelf, they should be ultra-low power themselves.

\subsubsection{\textbf{The ADC}} For reducing the ADC power consumption, a low-power ADC should be implemented in the application.

\subsubsection{\textbf{The Processor \& The Memory}} For the implementation of the BLE stack, a high performance processor is required. This processor, a power hungry one, is not rational to be enabled in the sleep intervals. On the other hand, a light real-time process on the data for packing and arranging the data packages in the sleep intervals is indispensable. In order to perform an optimized process, a co-processor is required for the mentioned light process during sleep intervals. Moreover, activating the huge memory of the high performance processor is not applicable during sleep intervals for designing an ultra-low power system. In contrast, the small memory of the co-processor is sufficient for collecting ECG data during sleep intervals. As a result, the mentioned co-processor is all that is needed for processing data packages during sleep intervals.

\section{ECG Hardware Design \& Implementation}

\begin{figure}[h]
    \centering   
    \includegraphics[width=0.45\textwidth]{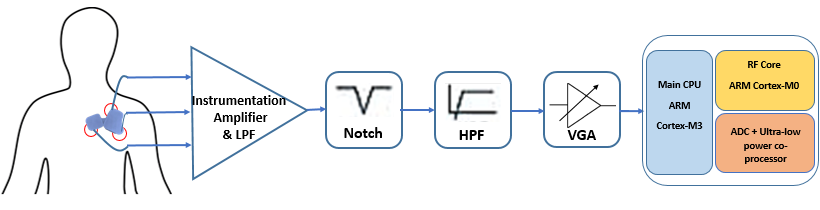}
    \captionsetup{justification=centering}
    \centering {\caption{\label{fig:3}Block diagram of the analog and digital circuits}}
\end{figure}

As discussed before, one of the main goals of this project is to design a device which makes every day self-monitoring of heart activities possible for individuals, so each part of the system should be cost-effective, ultra-low power and minimum sized. The block diagram of the analog and digital circuits is shown in Fig. \ref{fig:3} .
In the analog section, ultra-low power amplifiers with a maximum supply current of 1 $\mu $A have been used to minimize power consumption of the analog circuit. Likewise the core of the digital circuit, cc2650 Texas Instruments ultra-low power MCU for BLE, as the circuit\textquotesingle s main energy consuming block is programmed to operate in its lowest power consuming mode, which will be discussed in detail later. The cc2650 device contains a 32-bit ARM Cortex-M3 as the main processor, a unique ultra-low power sensor controller running on a separate 16-bit ARM Cortex-M0 processor and a 12-bit ADC with an ultra-low power analog comparator. \cite{DatasheetCC} 

The last block of the system uses LTC4054 (Lithium-ion battery charger from Linear Technology) to recharge the battery directly from USB port. The charger is chosen to meet the system\textquotesingle s power requirements in a way that its leakage current is sufficiently less than the total current consumption of the previous blocks. 

\subsection{Analog Front-End }  
The analog front end consists of filtering and amplifying stages designed to elevate the ECG signal and extract it from other in-band frequency components using various stages listed below:

\subsubsection{\textbf{Instrumentation Amplifier (IA)} }

The instrumentation amplifier as the first stage of the analog front-end contains two internal stages itself using ultra-low power amplifiers. As discussed before, the CMRR and input impedance of this stage meets ECG signal requirements. The IA stage is designed in a way that its programmable CMRR can reach beyond the maximum CMRR of the individual amplifiers to sufficiently diminish the common-mode effects of the user\textquotesingle s motions and surrounding interfering signals. 
The total CMRR for the IA is: 
  
\[CMRR_{total_{db}}= G1_{db}+CMRR_{Stage2_{db}}\addtag\] 
Where G1 is the differential gain of the first internal stage of the IA. The achieved CMRR  of this topology can reach up to 90 dB ,which is sufficient for our application. 

Even though the signal in the IA is extremely small, it is important to consider the large signal response of the circuit to analyze motion artifacts. Besides, there are some unexpected cases in wearable applications like lead off or electrode reattachment which result in a temporary large signal input. Since these artifacts appear like large signal inputs to the system, they can saturate the IA. 

To address this issue, \cite{Singapore} has introduced a baseline stabilizer which compares the differential input voltages with the normal common mode voltage of the circuit. In case of a large difference, two switches will be closed to turn the gain of the amplifier to one. This results in a quick return to the normal state. 

\begin{figure}[t]
    \centering   
    \includegraphics[width=0.45\textwidth]{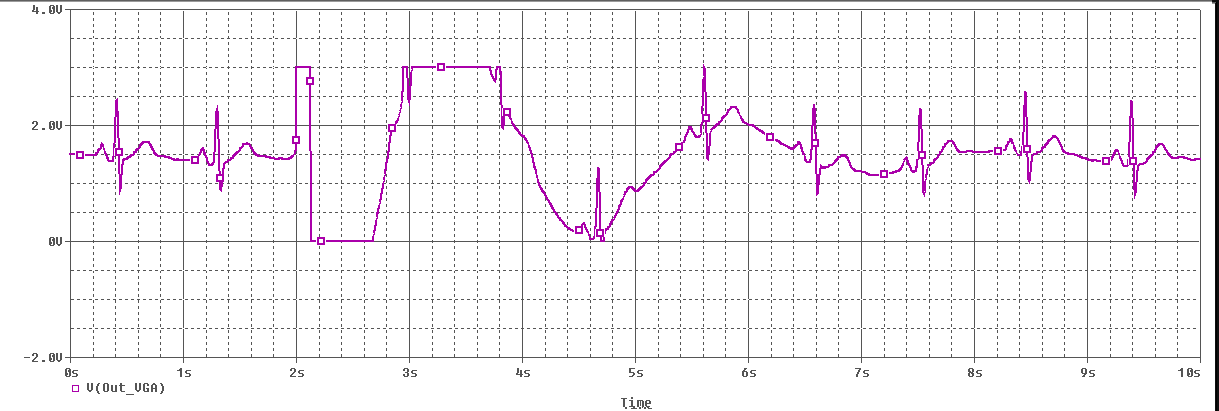}
    \captionsetup{justification=centering}
    \centering {\caption{\label{fig:4}Distorted ECG signal due to large signal interferences}}
\end{figure}

\begin{figure}[t]
    \centering   
    \includegraphics[width=0.45\textwidth]{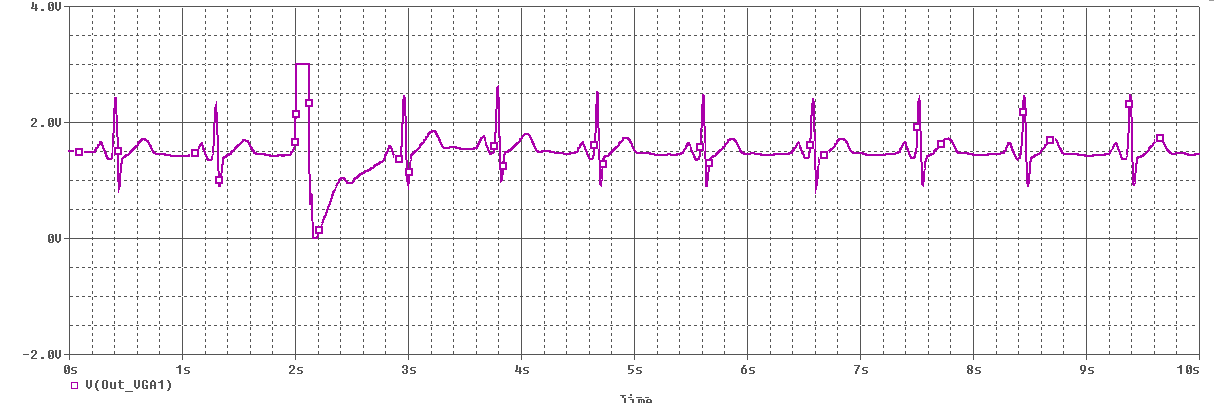}
    \captionsetup{justification=centering}
    \centering {\caption{\label{fig:5}ECG signal of our circuit with a decreased large signal settling time}}
\end{figure}

 In this work, because power consumption is of great concern, the operational amplifiers are chosen so that their power consumption is limited to 1 $\mu$A. Limiting the current of the analog circuit results in low slew rate and problems of large signal performance of the system. This may lead to the loss of a considerable proportion of the ECG signal (shown in Fig. \ref{fig:4}) unless the recovery time of the circuit is limited to a short period. Given these considerations, the IA is designed such that the large signal settling time is limited to a single heart cycle. As it is shown in Fig. \ref{fig:5}, the baseline is returned to its primary state in just a cycle of ECG signal.

Another consideration for the first stage is the signal path from patient\textquotesingle s body through skin-electrode contacts and the electrodes themselves which create a variable high impedance path from the body to the input of the instrumentation amplifier that can reach up to 0.62 M$\Omega $ \cite{AAMI_EC11}. The input impedance of the IA is high enough to ‏compensate for skin and electrode impedance deviations. The average skin and electrode impedance for Ag-Cl chest leads is 354 k$\Omega $ but can decrease to 20 k$\Omega $ after skin abrasion \cite{ProperSkin}. In this work, despite these impedance deviations, the trace quality has been preserved because of the high 22 M$\Omega $ input impedance of the designed IA. According to the American Health Association\textquotesingle s (AHA) report on recommendations for Electrocardiography, the cut-off frequency of the low pass filters should be equal to or higher than 150 Hz. First-order low pass filters with a cut-off frequency of about 150 Hz have been used in this stage to eliminate power line harmonics and adjacent high frequency components such as electromyogram signals.

\subsubsection{\textbf{Notch Filter}} 

The fundamental harmonic of the power line lies in the middle of the frequency band of ECG signal. Furthermore, compared to the weak microvolt ECG signal, power line signals with amplitudes of about 1.5 volt can substantially spoil the desired output of the system.  Instead of passive notch filters with limited attenuation, active notches with high quality factors have been used to attenuate the intervening power line component more than 86 dB while preserving adjacent in band frequencies like 30 Hz components with an attenuation less than 3dB.  

\subsubsection{\textbf{High pass Filter}}

The signals mentioned above are high passed at a cut-off frequency of 0.05 Hz to 1 Hz in order to suppress baseline wander. Extending the cutoff frequency helps rejecting baseline and DC offsets at the cost of ECG signal corruption because of the attenuated low frequency components of the main signal.

\subsubsection{\textbf{Variable Gain Stage}}
The gain of the circuit should be adjusted to cover the dynamic range of the 12-bit ADC for input signals with different amplitudes. The signal is grounded to 1.5 volt to cover the full 0 to VDD operating range of the ADC. The gain has been set to be about 68 dB depending on the input referred noise of the system and the dynamic range of the ADC. The overall system has an input referred noise of 0.49 $\mu V_{rms}$.

\subsection{Digital Circuit}

\begin{figure}[t]
    \centering   
    \includegraphics[width=0.45\textwidth]{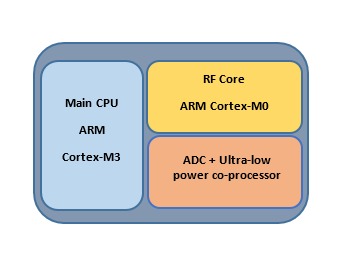}
    \captionsetup{justification=centering}
    \centering {\caption{\label{fig:6}The functional block diagram of cc2650}}
\end{figure}

The functional block diagram of the cc2650 device is shown in Fig. \ref{fig:6}.  The cc2650 device is a triple processor device exploiting Bluetooth Low Energy. The main CPU is a 32-bit ARM Cortex M3 processor running at 48 MHz with a unique ultra-low power sensor controller. The ARM Cortex-M3 CPU runs the application and higher levels of the protocol stack during the active mode with a normal operation of the processor and all of the peripherals that are currently enabled. This high performance processor must be turned off when it is not needed because in the on-state, it consumes a large amount of power.

The Bluetooth Low Energy controller and the IEEE 802.15.4 MAC are partly running on the RF core of the device which consists of an ARM Cortex-M0 processor. This processor is not programmable by the user.
The third part of the system is a co-processor which optimizes the power consumption of the digital circuit. It can collect analog and digital data autonomously while the rest of the system is in sleep mode. 

\begin{figure}[t]
    \centering   
    \includegraphics[width=0.45\textwidth]{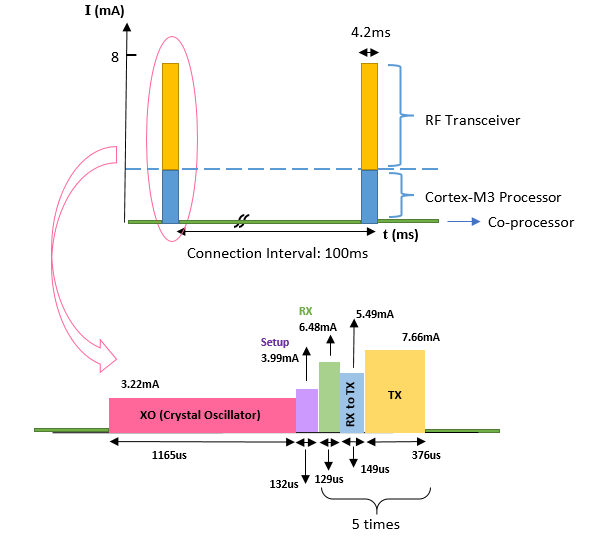}
    \captionsetup{justification=centering}
    \centering {\caption{\label{fig:7}Power consumption of cc2650 in each connection interval}}
    \end{figure}
    
The importance of the co-processor is that it enables the cc2650 device to collect data while the main M3 processor is turned off. Without the co-processor the digital circuit would consume a constant current of 3 mA because of the superfluous use of the main CPU. Fig. \ref{fig:7} shows the current consumption of the cc2650 device during each connection interval. The dashed lines indicate how much current the digital circuit would consume without having the co-processor optimize the power consumption. In active intervals when the main CPU is on, the main CPU and the RF core consume a current of 9 mA. The detailed power consumption of each packet during active intervals is shown in Fig. \ref{fig:7} \cite{Powerref}. After the active interval, power consumption of the device decreases to only 50 $\mu $A, because the co-processor can perform the tasks autonomously without the main CPU operation in the rest of the connection interval. The overall power consumption of the digital circuit is calculated as in:
\begin{equation}
\begin{array}{ll}
       \displaystyle I_{avg} = \frac{I_{RX}T_{RX}+I_{TX}T_{TX}+I_{XO}T_{XO}}{T_{interval}}\\ \\
      \displaystyle + \hspace{1mm} \frac{I_{setup}T_{setup}+I_{trans}T_{trans}}{T_{interval}} =
      \\ \\
       \displaystyle  + \hspace{1mm} \frac{6.48 mA \times 129 \mu s + 7.66 mA \times 1880 \mu s }{100 ms}
      \\ \\
       \displaystyle  + \hspace{1mm} \frac{3.22 mA \times 1165 \mu s + 3.99 mA \times 132 \mu s}{100 ms}
       \\ \\
        \displaystyle  + \hspace{1mm} \frac{5.49 mA \times 149 \mu s \times 6}{100 ms} = 244 \mu A

\end{array}
\end{equation}

In standby mode when the main processor is turned off, an external wake event such as a co-processor event is required to bring the device from standby mode back to active mode and the CPU continues execution from where it went into standby mode.
The analog front end (AFE) extracts and preprocesses the ECG signal so that it can cover the input range of the ADC of the sensor controller. The Sensor Controller contains circuitry that can be selectively enabled in standby mode including the 12-bit, 200 Ksps ADC.
The converted signal is transmitted in packets of 20 bytes through the RF link, Bluetooth Low Energy, to the Bluetooth receiver of the user\textquotesingle s smartphone.
\cite{DatasheetCC}

The analog and digital sections have been mounted on two pieces of double layer PCB boards which can be easily connected to the body using three button electrodes shown in Fig. \ref{fig:1}. Dividing the circuit into two segments improves the device's robustness against motion artifacts by enabling relative movements of the probes. 

\section{User interface}
In order to present a whole system in today\textquotesingle s biomedical device market, providing a qualified, user-friendly software is of great significance. In this project, the software includes three synchronously connected parts. These three parts together, link the patients and the physicians, and provide them with a real-time remote health monitoring system. There is a description on each part bellow. 

\subsection{Smartphone application} 

As mentioned before, the signal recorded by the sensor will then be transmitted to smart phone via BLE. There is a need for a well-designed application on smart phone to handle the received data. In each data packet, a byte is allocated to an incrementing sequence. By this means, the mentioned application on the smart phone can detect any packet loss by monitoring the sequence. In case of a packet loss, a feedback loop is implemented in order to retrieve the lost packet. Therefore, this application is reliable enough not to lose a single data packet. After being received on the smart phone, the data packets will be stored in a phone-oriented database. The application is capable to visualize the patient\textquotesingle s ECG line real-time. It has the feature to note the physical and emotional mode of the patient while recording data or any desired note about a certain time period of registered signal to later inform the physician. Last but not least, the application prepares the data for the next wireless transmissions to patient\textquotesingle s personal computer or to the cloud for physician’s access. 

\begin{figure}[t]
    \centering   
    \includegraphics[width=0.45\textwidth]{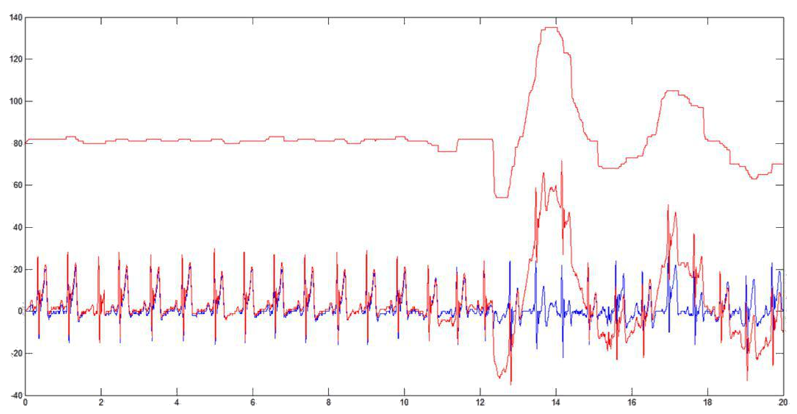}
    \captionsetup{justification=centering}
    \centering {\caption{\label{fig:8}Motion artifact removal in the smartphone application}}
\end{figure}

Moreover, the application operates signal processing on the received ECG signal. It looks for suspicious traces such as those with missing pulses, tachycardia, bradycardia, etc. In case of finding suspicious traces, the application marks the related portion of the signal to be further analyzed by the physician. The signal processing is also concerned with motion artifact removal. As explained before, the common-mode motion artifact is rejected by the high CMRR of the analog front end. However, there are in-band differential motion artifacts which cannot be handled by the analog front end. In the smart phone application, these in-band differential motion artifacts are rejected programmatically by digital signal processing. Figure \ref{fig:8} shows the strong removal of motion artifacts done by the smart phone application.

\subsection{Personal computer application} 

Due to the limited memory of a smartphone and its relatively small screen size, preparing a computer application synchrnously connected to the smartphone is so beneficial. Besides, a computer has better data analyzing capabilities compared to a smartphone. So the second software part is an application on the patient’s personal computer receiving data from smartphone via Wi-Fi, without any need to the internet. It stores the received data efficiently on a database as backup. The application is able to show the ECG lines of any time period with any length in a proper resolution.

\subsection{Physician\textquotesingle s dashboard} 
 
In pursuing the target of “remote” health monitoring, considering an access to the recorded data for the corresponding physicain is a relevant concern. The third software part is the physician\textquotesingle s dashboard. The smartphone application uploads the data on cloud. Physician\textquotesingle s dashboard has access to this cloud and to the recorded data of their patients. So the physician is able to monitor the patient in real-time. The dashboard sorts the patients on the basis of their state's emergency and shows the patients\textquotesingle emergency alarms as notifications. Dashboard can also contain the patients\textquotesingle medical history.

\section{Comparison with similar works}
In this section, a comparison between our work and some related works is presented. As shown in table I, this work exceeds the AHA specification by a significant margin in most parameters.

\begin{table}[t]
  \centering
  \caption{AHA specification comparison.}
    \begin{tabular}{|c|c|c|}
    \hline
    Parameter & AHA standard & This work \\ \hline
    ADC ENOB & $>$ 8-10 bits & 12 bits \\ \hline
    ADC sample rate & $>$ 120 Hz & 1000 bits \\ \hline
    Leads & From 1 to 12 & 1 \\ \hline
    Input noise & $<$ 20$\mu$Vrms  & 0.49 $\mu$Vrms \\ \hline
    CMRR  & $>$ 60dBm & 90 dBm \\ \hline
    F min & 0.05 – 0.5 Hz & 0.05 Hz \\ \hline
    F max & 40 – 100 kHz & 150 Hz \\ \hline
    lifetime & $>$ 1 - 2 days & 20 days \\ \hline
    \end{tabular}
  \label{tab:addlabel}
\end{table}
 
Given the diversification of the recent mobile ECG sensors, only the similar works with the following criteria are chosen, in order to be compared with our own work \cite{Jsen}: 
·         Battery-based wearable sensors with an operation time more than 24 hours 
·         High quality signals with at least 10 bits, a sampling frequency of at least 100 hz, and a CMRR of at least 60 dBm) 
·         Whole sensors (containing AFE, ADC, MCU, power management, and radio) 
Table II presents the mentioned sensors (\cite{JsenTwentyTwo}, \cite{JsenTwentyThree}, \cite{JsenFourtyFive}, \cite{JsenTen}, \cite{JsenFourtyNine}, \cite{JsenFourtySix}, \cite{Jsen}) and the data about the most important parameters. 
The most discernible fact represented in table II is that this work has an ultra-low power consumption in comparison with the similar works with different communication protocols, thanks to the use of the co-processor which is explained in section IV. This work has a power consumption of less than 3 mW, while the minimum power consumption among other works is 12 mW which belongs to \cite{Jsen}. 
Furthermore, our ECG sensor enjoys a significantly longer lifetime than the related works. The longest lifetime among the other works is 160 hours \cite{Jsen} which is about one third of this work\textquotesingle s longevity. 
While many of the mentioned works use a driven right-leg circuit, we do not use it since it is neither necessary nor convenient, as is explained in the third section.

\begin{table*}[t!]
\small
  \centering
  \caption{Performance comparison table.}
    \resizebox{\linewidth}{!}{
    \begin{tabular}{|c|c|c|c|c|c|c|c|}
    \hline
          & This work  & \cite{JsenTwentyTwo}-\cite{JsenTwentyThree} & \cite{JsenFourtyFive}  & \cite{JsenTen}  & \cite{JsenFourtyNine}  & \cite{JsenFourtySix}  & \cite{Jsen} \\ \hline
    ADC (No. of bits)  & 12    & 12    & 24    & 12    & 10    & 14    & 16.5 \\ \hline
     (Hz) & 1000  & 256   & 160   & 512   & 500   & 100   & 320 \\ \hline
     (Hz) & 0.05  & n/a   & n/a   & 0.05  & 1     & 0.8   & 0.1 \\ \hline
     (Hz) & 150   & n/a   & n/a   & 150   & 150   & n/a   & 153 \\ \hline
    CMRR (dB) & 90    & $>$100  & n/a   & n/a   & 50    & n/a   & $>$112 \\ \hline
    Input noise (nVRMS) & 490   & $<$1000 & 1500  & n/a   & n/a   & n/a   & 480 \\ \hline
    DRL circuit & No    & no    & Yes   & Yes   & Yes   & n/a   & No \\ \hline
    Communication protocol & BLE   & BLE   & BLE   & BT    & BT    & Modified ZigBee & ZigBee Pro \\ \hline
    Packet loss & No packet loss & n/a   & n/a   & n/a   & n/a   & 10$\%$ (5 nodes), 4$\%$ (1 node) & $<$0.025$\%$ (1 node), $<$6$\%$ (6 nodes) \\ \hline
  
    Voltage (V) & 3     & 3.7   & 3.6   & 3.7   & 3.7   & 4.1   & 3 \\ \hline
    I Avg. (mA) & $<$0.3  & 5.8   & 15    & 31    & 20    & 41.8  & 4.07 \\ \hline
    Power (mW) & $<$0.9  & 21.48 & 54    & 115   & 74    & 171.4 & 12 \\ \hline
    Battery (mAh) & 150   & 400   & 720   & 1100  & 280   & 500   & 650 \\ \hline
    Life (hour) & $>$480  & 70    & n/a   & 33    & 24    & 10    & 160 \\ \hline
    \end{tabular}}
  \label{tab3}
\end{table*}

A quality that distincts our work from all other devices is the real-time monitoring feature which enables the physician to be remotely aware of one\textquotesingle s patients\textquotesingle status in any desired time. In fact, none of the works mentioned in table II provides such a possibility for both patients and physicians to be connected during day and night, in case of emergencies, or even for tracking daily ECG disorders.

\section{Measurement and Experimental Results}
\begin{figure}[t]
    \centering   
    \includegraphics[width=0.45\textwidth]{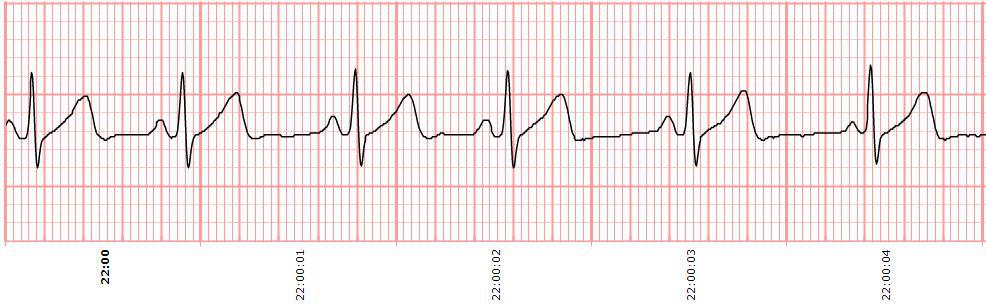}
    \captionsetup{justification=centering}
    \centering {\caption{\label{fig:9}ECG signal resulting from our system}}
\end{figure}

The final result of this work is shown in Fig. \ref{fig:9} which is a demonstration of ECG signals in the user\textquotesingle s smartphone application. The common-mode noises and power line interference are eliminated and the original ECG signal is reinforced in gain stages to cover the 0 to 3 v range of the ADC. The implemented ADC in cc2650 device samples the signal at the input of the digital circuit with a sampling rate of 1000 sps.

\begin{figure}[t]
    \centering   
    \includegraphics[width=0.45\textwidth]{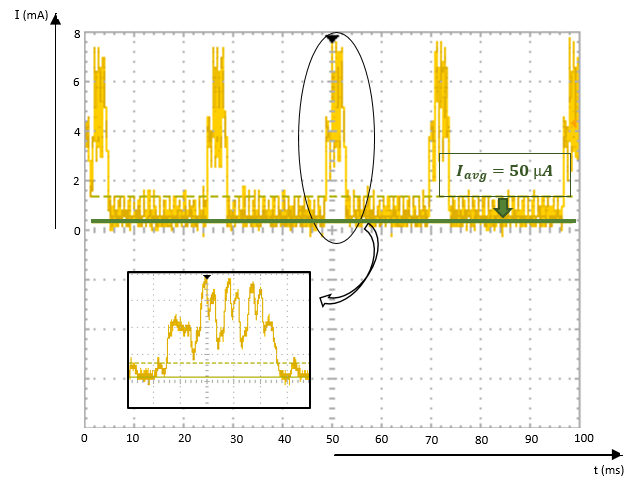}
    \captionsetup{justification=centering}
    \centering {\caption{\label{fig:10}Power consumption of the digital circuit}}
\end{figure}

In order to verify equ.4, power consumption of the digital circuit are measured and the results are depicted in Fig. \ref{fig:10}.

\section{Conclusion and Future Work}
The proposed ultra-low power ECG monitoring system provides a convenient everyday monitoring both for the patients and their physicians. This robust system enables the users to track their hearts\textquotesingle signals even during intensive physical activities by removing the motion artifacts and filtering the external noises. The system utilizes BLE as an ultra-low power RF link to transmit ECG data to the user\textquotesingle s smartphone in the form of packets. In addition to the possibility of a real-time self-monitoring on the smartphone, patients can add their physical moods while recording their ECG data to better inform the physician about their medical status. The personal computer application and the physician\textquotesingle s dashboard facilitate additional monitoring and processing of the ECG data for the patient and the physician respectively. The mentioned user interfaces together with the robust ultra-low power sensor circuitry provide heart patients and even healthy people with a comfortable device to keep track of their hearts\textquotesingle activities and enables physicians to come up with more practical and prompt treatments. 
The future work can focus on improving signal processing to detect arrhythmia in user\textquotesingle s smartphone and in the personal computer application even before sending the data to the physician. Extra work can be done to send notifications to the emergency contacts and to emergency agencies under medical crises.

\ifCLASSOPTIONcaptionsoff
  \newpage
\fi

\section*{Acknowledgment} 

The  authors  would  like  to  thank  Alireza Amirshahi and  Azam Mokhtari for their generous contribution in this project.

\vspace{-4em}
\begin{IEEEbiography}[{\includegraphics[width=1in,height=1.25in,clip,keepaspectratio]{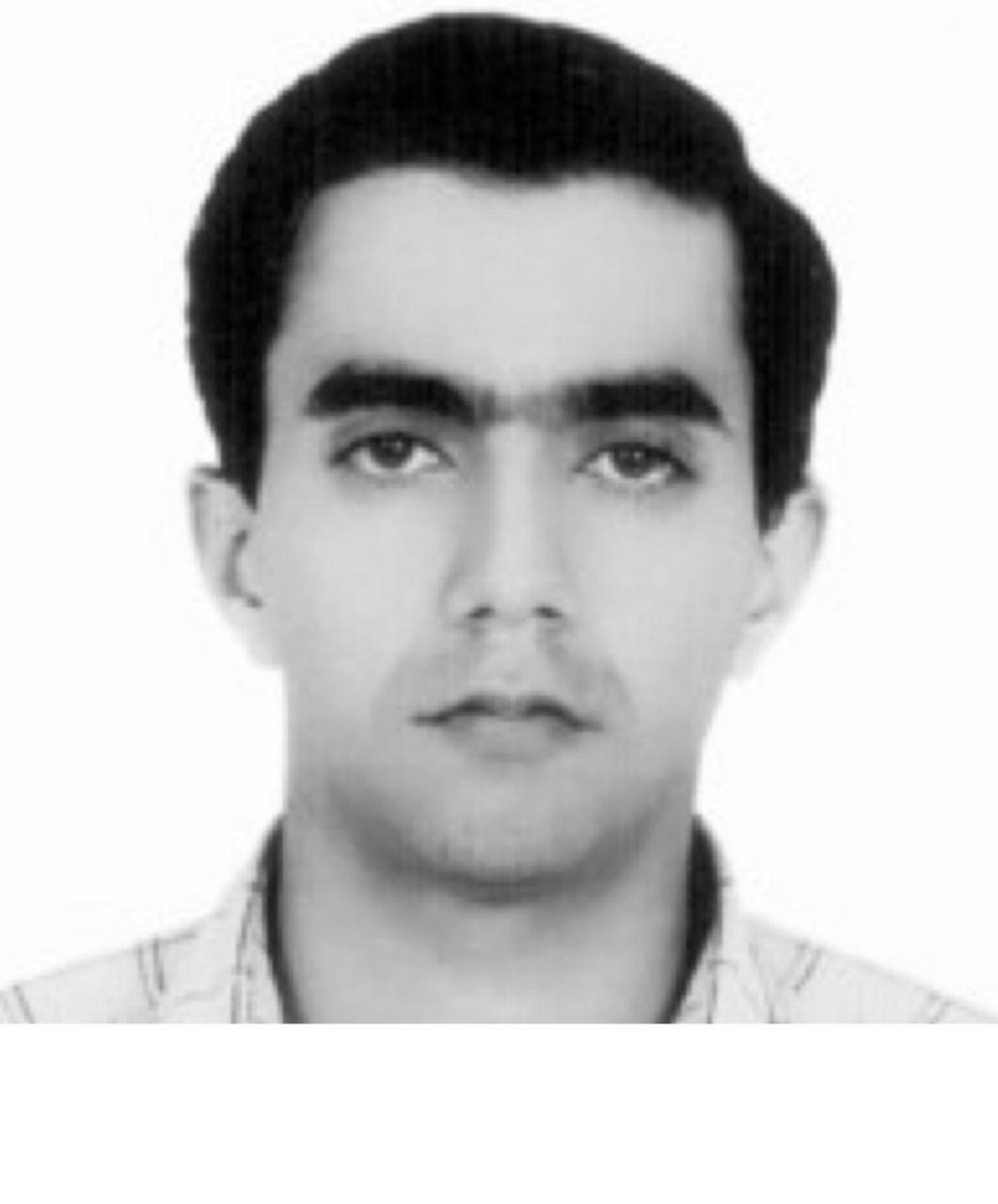}}]{Ehsan Hadizadeh}
Ehsan Hadizadeh hafshejani was born in Shahrekord, Iran in 1989. He received a B.S. degree in Electrical Engineering from Shahrekord University in 2010 and a M.S. degree from Sharif University of Technology in 2013. He is currently a Ph.D. student at Sharif University of Technology and a visiting research scholar at the Electrical and Computer Engineering department of the University of British Columbia. His research interests include RF/Analog IC design and ultra-low power systems.
\end{IEEEbiography}
\vspace{-4em}
\begin{IEEEbiography}[{\includegraphics[width=1in,height=1.25in,clip,keepaspectratio]{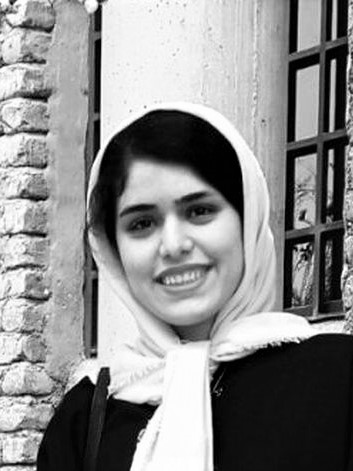}}]{Rozhan Rabbani}
Rozhan Rabbani was born in Urmia, Iran in 1995. She received he B.S. degree in Electrical Engineering from Sharif University of Technology. She is currently a Ph.D. student in Electrical Engineering at University of California, Berkeley. Her research interests include Analog and Mixed signal IC design, Biomedical circuits and sensors, Energy Harvesting and Low power system design and Optoelectronics.
\end{IEEEbiography}
\vspace{-4em}
\begin{IEEEbiography}[{\includegraphics[width=1in,height=1.25in,clip,keepaspectratio]{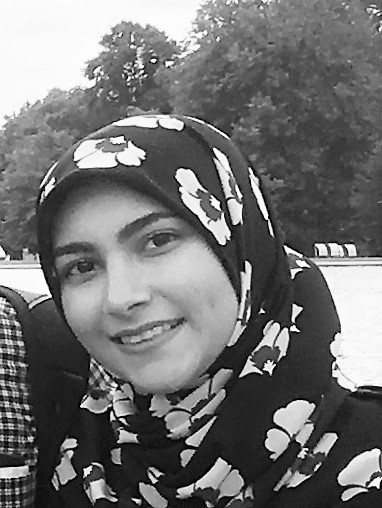}}]{Zohreh Azizi}
Zohreh Azizi was born in Tehran, Iran in 1994. She received her B.S. degree in Electrical Engineering from Sharif University of Technology in 2018 and joined the Ph.D. program in Electrical Engineering, electrophysics at the University of Southern California. She is currently a Research assistant at the University of Southern California. Her research interests include MEMS transducers, piezoelectric thin-film devices, hardware design in Biomedical applications , AI and signal processing, and electrical measurement.
\end{IEEEbiography}
\vspace{-4em}
\begin{IEEEbiography}[{\includegraphics[width=1in,height=1.25in,clip,keepaspectratio]{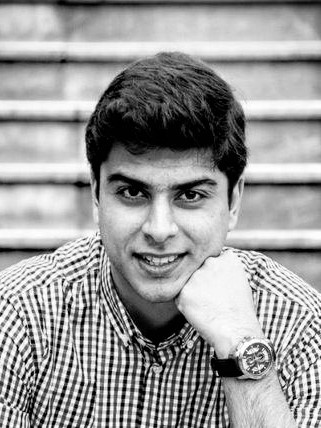}}]{Matin Barekatain}
Matin Barekatain was born in Esfahan, Iran in 1995. He received his B.S. degree in Electrical Engineering from Sharif University of Technology in 2017 and joined the Ph.D. program in Electrical Engineering, electrophysics at the University of Southern California. He is currently a Research assistant at the University of Southern California. His research interests include MEMS transducers, piezoelectric thin-film devices, hardware design, AI and signal processing, and electrical measurement.
\end{IEEEbiography}
\vspace{-4em}
\begin{IEEEbiography}[{\includegraphics[width=1in,height=1.25in,clip,keepaspectratio]{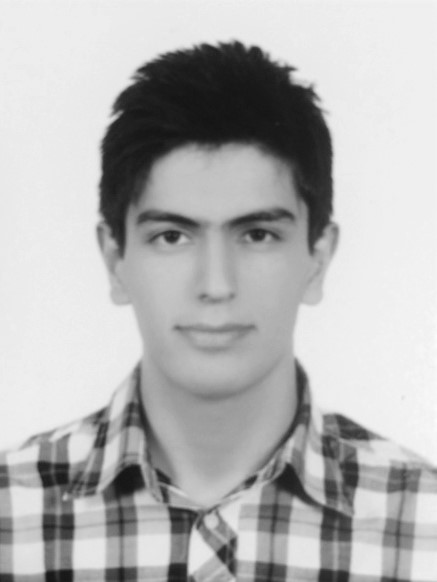}}]{Kourosh Hakhamaneshi}
Kourosh Hakhamaneshi was born in Tehran, Iran in 1995. He earned his B.S. degree in Electrical Engineering at Sharif University of Technology. He is currently pursuing his Ph.D. program in EECS department at University of California Berkeley. His research focuses on energy-efficient deep learning systems on edge, at the intersection between machine learning and computer architecture.
\end{IEEEbiography}
\vspace{-4em}
\begin{IEEEbiography}[{\includegraphics[width=1in,height=1.25in,clip,keepaspectratio]{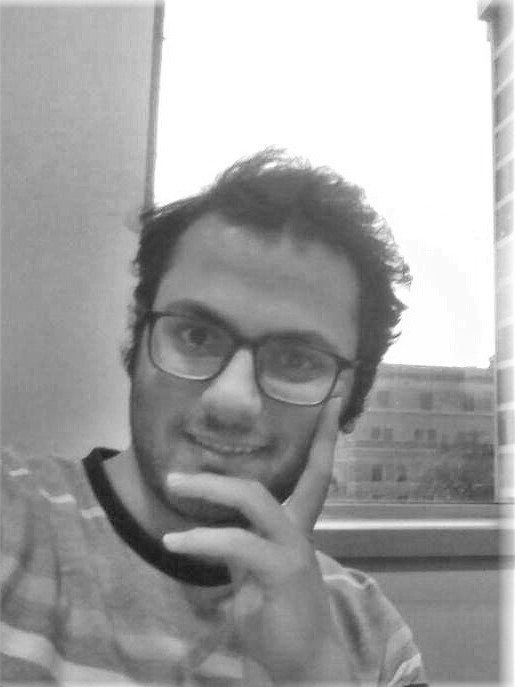}}]{Erfan Khoram}
Erfan Khoram was born in Tehran, Iran in 1994. He Received his B.S. degree in Electrical Engineering from Sharif University of Technology in 2017, and Joined University of Wisconsin-Madison to pursue his Ph.D. in ELectrical Engineering. He is currently a research assistant at University of Wisconsin-Madison. His current research interests include depth sensing, and application and implementation of Neural Networks in photonic systems.
\end{IEEEbiography}
\vspace{-4em}
\begin{IEEEbiography}[{\includegraphics[width=1in,height=1.25in,clip,keepaspectratio]{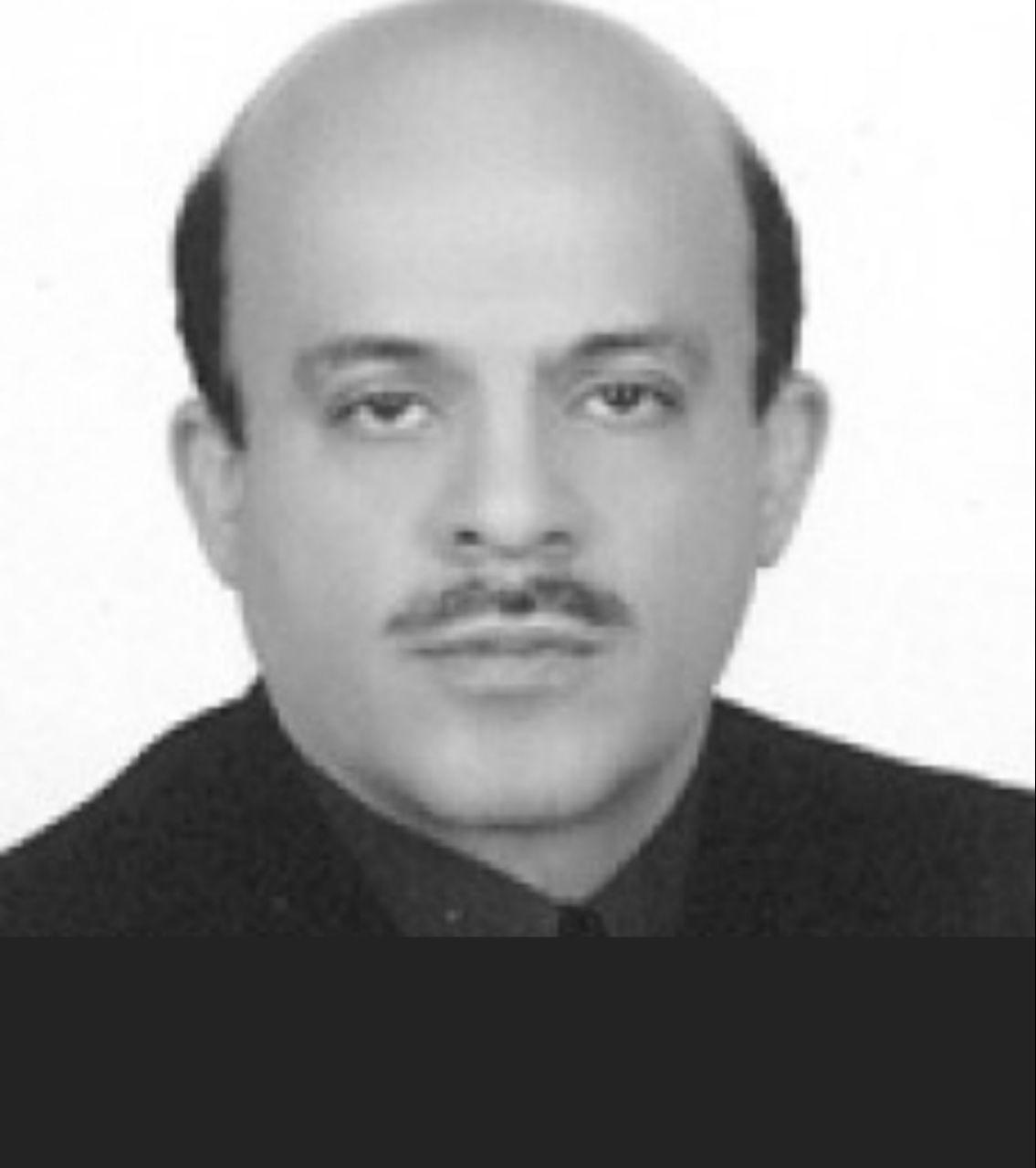}}]{Ali Fotowat-Ahmady}
Ali Fotowat-Ahmady was born in Tehran, Iran, in 1958. He received the B.S. degree from the California Institute of Technology, Pasadena, in 1980, and the M.S. and Ph.D. degrees in electrical engineering from Stanford University, Stanford, CA, in 1982 and 1991, respectively.
He started his career at Philips Semiconductor in Sunnyvale, CA, in 1987, where he developed several integrated circuits for mobile phones. In 1991, he joined the Department of Electrical Engineering, Sharif University of Technology, Tehran. His research interests include advanced integrated circuits for energy savings and communication/positioning applications. Due to his interests in entrepreneurial engineering, he has been the co-founder of several companies and continues advising his students on the same.
Dr. Fotowat-Ahmady is a member of the IEEE Solid State Society and has been the adviser of the society\textquotesingle s Sharif EE Student Chapter. He is a three times recipient of Kharazmi Science and Engineering Award for his work on lowpower microelectronics and communication IC\textquotesingle s.
\end{IEEEbiography}

\end{document}